\documentclass[5p,preprint,sort&compress]{elsarticle}
\pdfoutput=1
\usepackage{amsmath}
\usepackage{amssymb}
\usepackage{fancyhdr}

\usepackage{wrapfig}
\usepackage{xspace}
\usepackage{pstricks}
\usepackage{graphicx}
\usepackage{setspace}
\usepackage{threeparttable}
\usepackage{relsize}
\usepackage{ifthen}
\usepackage{lineno}
\usepackage{enumerate}
\usepackage{makeidx}
\usepackage{slashed} 
\usepackage{multirow}
\usepackage{placeins}
\usepackage{float}

\newcommand{\order}{\ensuremath{\mathcal{O}}}

\newcommand{\bea}{\begin{align}}
\newcommand{\eea}{\end{align}}
\newcommand{\beq}{\begin{equation}}
\newcommand{\eeq}{\end{equation}}

\newcommand{\bs}{\begin{split}}
\newcommand{\es}{\end{split}}

\newcommand{\bi}{\begin{itemize}}
\newcommand{\ei}{\end{itemize}}
\newcommand{\bc}{\begin{center}}
\newcommand{\ec}{\end{center}}
\newcommand{\bac}{\begin{array}{c}}
\newcommand{\bacc}{\begin{array}{cc}}
\newcommand{\ea}{\end{array}}

\def\spa#1.#2{\langle#1\,#2\rangle}
\def\spb#1.#2{[#1\,#2]}

\newcommand{\Ummeter}{\ensuremath{\mathrm{mm}}\xspace}

\newcommand{\UMeV}{\ensuremath{\mathrm{MeV}}\xspace}
\newcommand{\UGeV}{\ensuremath{\mathrm{GeV}}\xspace}
\newcommand{\UTeV}{\ensuremath{\mathrm{TeV}}\xspace}

\newcommand{\sla}[1]{\ensuremath{{#1\kern-0.45em/}}}

\newcommand{\Sherpa}{S\protect\scalebox{0.8}{HERPA}\xspace}

\newcommand{\Comix}{C\protect\scalebox{0.8}{OMIX}\xspace}

\newcommand{\CSShower}{CSS\protect\scalebox{0.8}{HOWER}\xspace}

\newcommand{\Ahadic}{A\protect\scalebox{0.8}{HADIC\plusplus}\xspace}

\newcommand{\Photons}{P\protect\scalebox{0.8}{HOTONS\plusplus}\xspace}

\newcommand{\Rivet}{R\protect\scalebox{0.8}{IVET}\xspace}

\DeclareRobustCommand{\plusplus}{\raisebox{0.2ex}{\smaller++}}

\newcommand\lapproxeq{\lower .7ex\hbox{$\;\stackrel{\textstyle <}{\sim}\;$}}
\newcommand\gapproxeq{\lower .7ex\hbox{$\;\stackrel{\textstyle >}{\sim}\;$}}

\newcommand\cls{\ensuremath{\mathrm{CL_{s}}}\xspace}

\graphicspath{{figures/}}
\date{}

\title{\vspace*{-2cm}{\scriptsize  \mbox{}\hfill IPPP/22/07\\[0.5cm]}}
\title{Constraining the Charm-Yukawa coupling at the Large Hadron Collider}
\author[1,2]{Joseph Walker}
\author[1,2]{Frank Krauss}
\address[1]{Institute for Particle Physics Phenomenology, Durham University, United Kingdom}
\address[2]{Institute for Data Science, Durham University, United Kingdom}

\begin{document}
\begin{abstract}
We present theoretical results for the sensitivity of charm Yukawa coupling measurements in future high-luminosity LHC runs in three channels: Vector Boson Fusion (VBF), $W$ Higgs-strahlung and $Z$ Higgs-strahlung production of a Higgs boson and its subsequent decay into charm quarks.
To reduce the overwhelmingly large backgrounds and to reduce false positives, we apply a set of simple kinematic and jet feature cuts and feed neural network data structures of three types; jet features, jet images and particle level features.
To facilitate straightforward comparison with experimental studies~\cite{ATLAS:2021zwx, CMS:2019tbh}, we express our results in terms of signal strengths~\cite{lhchiggscrosssectionworkinggroup2012lhc}.
\end{abstract}
\maketitle

\section{Introduction}

The 2012 discovery of the Higgs boson $H$ with a mass of $125\UGeV$ by the ATLAS and CMS collaborations at the LHC~\cite{Aad_2012,CMS:2012qbp} completed the Standard Model (SM) of particle physics, and initiated a step-change in our quest to understand nature at its fundamental scale.
Since this discovery the focus has shifted to measuring the new boson's properties and interactions, and to constrain effects of possible new physics manifesting itself through subtle deviations from SM predictions.
A central part of these efforts has been the determination of the Higgs boson couplings to the other SM particles.  
By the end of Run 2 couplings to the SM gauge bosons~\cite{10.1093/ptep/ptaa104,2019134949,Sirunyan:2765059} and to the massive third-\-generation fermions~\cite{10.1093/ptep/ptaa104,PhysLetsB:2018:2018173, Sirunyan:2743741, PhysRevD.99.072001, 2018283, 2021136204, Sirunyan:2636067} have been measured, and the first coupling to the second-generation fermions, i.e.\ to the muon, has been constrained~\cite{10.1093/ptep/ptaa104,Sirunyan:2730058, 2021135980}.

Taking advantage of the significantly higher luminosity of the forthcoming phases of LHC data taking, the precision of these measurements will be further improved and couplings that have hitherto been inaccessible will become subject to scrutiny.  
This includes, in particular, a determination of the Higgs Yu\-ka\-wa coupling to charm quarks as a further, stringent test of the universal role of the Higgs boson in the generation of fundamental masses.
We will analyse the prospects for such a measurement in three channels, namely the production of Higgs bosons and their subsequent decay into charm quarks in association with $Z$ or $W$ bosons and in vector boson fusion,
\begin{equation}
    \begin{array}{lcll}
        pp &\to & H_{\to cc}jj+X & 
        \mbox{\rm (0-lepton channel, 0L)} \\
        pp &\to & W_{\to \ell\nu_\ell}H_{\to c\bar c}+X & 
        \mbox{\rm (1-lepton channel, 1L)}\\
        pp &\to & Z_{\to \ell\bar\ell}H_{\to c\bar c}+X & 
        \mbox{\rm (2-lepton channel, 2L)}
        \nonumber
    \end{array}
\end{equation}
In all processes the relevant quantity is the branching ratio of the Higgs boson to the charm quark, in the SM given by $Br_{H\to c\bar{c}} = 2.89\%$~\cite{10.1093/ptep/ptaa104} and significantly smaller than its counterpart, the branching ratio to the $b$ quarks, around $58.2\%$~\cite{10.1093/ptep/ptaa104}.
In previous studies~\cite{ATLAS:2021zwx, CMS:2019tbh}, therefore, the production of $b$ quarks has been identified as a significant non-trivial background, along with signatures from QCD, vector boson and $t\bar{t}$ production. 

To provide a simple interpretation of our results we will express the signal strength $\mu$ in the $\kappa$-framework~\cite{lhchiggscrosssectionworkinggroup2012lhc} and use the \cls method~\cite{Read_2002} to estimate the upper bound for $\kappa_c$ at the 95\% level. 
Modifying our observed signal counts, $s_x$ by the signal strength $\mu_i$ in each bin, $x$ of a given distribution,
\begin{equation}
    N_x(\mu) = b_{x} + \mu s_{x}\,,
\label{signalstrength}
\end{equation}
and assuming an unmodified background, $\mu=1$ defines the SM value and varying is equivalent to varying $\kappa_c$ in some parametric way derived below. 
For different signal processes $i$, i.e.\ for the 0-lepton, 1-lepton, and 2-lepton signatures, the $\mu_i$ represent ratios of cross sections $\sigma_i$ times branching ratios in the narrow-\-width approximation:
\begin{equation}
    \mu_{i} = \dfrac{\sigma_{i} \cdot \textrm{Br}_{H\to c\bar{c}}}{\sigma_{i}^{\textrm{SM}} \cdot \textrm{Br}_{H\to c\bar{c}}^{\textrm{SM}}}\,,
\label{eq:Mu1}
\end{equation}
where the $\sigma_i$ denote the cross sections of the Higgs boson production processes $i$ and $Br_{H\to c\bar{c}}$ is the branching ratio of the Higgs boson to charm quarks, $\Gamma_{H\to c\bar{c}}/\Gamma_H$.  
The superscript ``SM" indicates the SM values, its absence refers to values obtained under variations of the signal strength.
In the $\kappa$ framework all couplings of particles $X$ to the Higgs boson are independently modified by multiplying them with some $\kappa_X$; here and in the following we will assume that only the charm Yukawa coupling is modified by a factor $\kappa_c$, which in turn will result in modified partial and total decay widths of the Higgs boson, and thus 
\begin{equation}
    \mu_{i} = 
        \left[\dfrac{\kappa_c^2\Gamma_{H\to c\bar{c}}^{\textrm{SM}}}{\Gamma_H^{\textrm{tot}}}\right]/
        \left[\dfrac{\Gamma_{H\to c\bar{c}}^{\textrm{SM}}}{\Gamma_H^{\textrm{tot, SM}}}\right]
\label{eq:Mu2}
\end{equation}

Current analysis with such a framework sets the limit at 95\% confidence at $\mu_{\textrm{VH}(c \bar{c})} \le 31^{+12}_{-8}$ at ATLAS~\cite{ATLAS:2021zwx} measured at an integrated luminosity of $139 \textrm{fb}^{-1}$ and $\mu_{\textrm{VH}(c \bar{c})} \le 37^{+16}_{-11})$ at CMS ~\cite{CMS:2019tbh} measured at $35.9 \textrm{fb}^{-1}$.
The quadratic dependence of $\kappa_c$ on $\mu$ leads to some limitations in the maximal resolvable value of $\mu$ that leads to a meaningful value of $\kappa_c$. This value sits at $\mu = 1/Br_{H\to c\bar{c}}^\textrm{SM} = 34.6$.
\section{Simulation}
For our analysis, signal and background samples are generated with \Sherpa 2.2.9~\cite{Bothmann:2019yzt}, using LO-merged samples throughout~\cite{Catani:2001cc,Krauss:2002up,Hoeche:2009rj}.
For all signal and background processes considered, namely;
\begin{itemize}
\item vector boson fusion (0L), 
\item $WH$-associated production (1L),
\item $ZH$-associated production (2L),
\item vector boson ($W\to \ell\nu$, $Z\to\ell\bar\ell$) production + jets,
\item vector-boson pair production + jets, where one of the two bosons decays hadronically,
\item pure QCD multijets,
\item $t\bar t$ production + jets,
\end{itemize}
we merge up to two additional jets to the core process. 
We use the NNPDF 3.0 PDF~\cite{NNPDF30} from LHAPDF~\cite{Buckley:2014ana}, the \Comix matrix element generator~\cite{Gleisberg:2008fv} for the LO matrix element, the \CSShower~\cite{Schumann:2007mg} for the simulation of QCD radiation, \Ahadic as hadronisation model~\cite{2004}, \Photons~\cite{Schonherr:2008av} for the emission of photons in the decays of the $W$ and $Z$ bosons, and \Sherpa's buily-in models for the underlying event and hadron decays.
Using the default prescription for setting the renormalisation and factorisation scales in multi-jet merging, we obtain theoretical uncertainties from their variation by a factor of $f_{R,F}=2$ in both directions and forming the envelope of the 7-points, schematically
\begin{equation}
 \{f_R,\,f_F\} = \left\{\frac12,\,\frac12\;; \frac12,\,1\;; 1,\,\frac12\;; 1,\,1\;;1,\,2\;;2,\,1\;;2,\,2\right\} \,.  
\end{equation}
The rationale for using this approach is two-fold:
First of all, including higher-order QCD corrections does not induce any sizable change in the shape of distributions, but only alters the overall cross section which can usually be captured by applying a flat $K$-factor to the overall sample.  
For the processes we consider, this $K$-factor is of the order of 1.3 or below, thereby increasing the total number of events by up to 30\%, which in turn translates to a decrease of the statistical uncertainty by about 10\%.
Secondly, apart from increasing total event numbers, the higher-order corrections reduce scale uncertainties, typically by a factor of 2 or more.
As we are only able to roughly estimate experimental uncertainties, our approach of potentially overestimating the theory uncertainties therefore merely translates into our results being more conservative.
\section{Analysis Strategy}
\subsection{Initial Cuts}
\begin{figure}[tbp]
\centering 
\includegraphics[width=\linewidth]{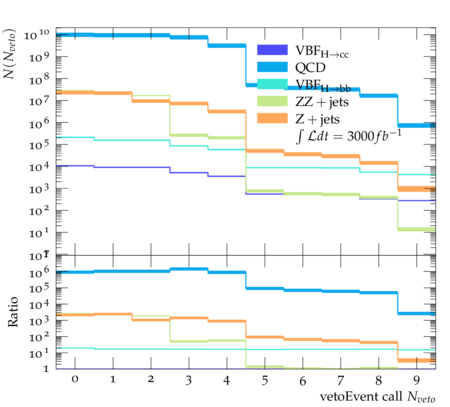}
\parbox{0.8\linewidth}{\caption{\label{fig:0LepCutflow} Example cut flow for the ``0L" channel. Shaded region indicates the error estimated from PDF4LHC scale variation convention and statistics. Backgrounds where $N_{\textrm{process}} <<  N_{\textrm{VBF}_{H\rightarrow c\bar{c}}}$ are omitted.}}
\end{figure}

\noindent
For each of the three signal topologies, there is a unique set of cuts, summarised in Tab.~\ref{tab:Cutflow}.
They are encoded in a \Rivet~\cite{Buckley:2010ar} analysis and detailed by:
\begin{table}
\begin{center}
 \begin{tabular}{ ||  c c || c c c || } 
 \hline
 Cut & \# & ``0L" & ``1L" & ``2L" \\ [0.5ex] 
 \hline\hline
 MET $\leq 30\,\UGeV$ & 1 & \checkmark & X & \checkmark \\
 \hline
 MET $\geq 30\,\UGeV$ & 1 & X & \checkmark & X \\
 \hline
 0 Isolated Leptons & 2 & \checkmark & X & X \\
 \hline
 1 Isolated Leptons & 2 & X & \checkmark & X \\
 \hline
 2 Isolated Leptons & 2 &X & X & \checkmark \\
 \hline
 1+ Fat jet & 3 & \checkmark & \checkmark & \checkmark \\
 \hline
 Candidate Fat jet  & 4 & \checkmark & \checkmark & \checkmark \\
 \hline
 2 forward QCD jets & 5 & \checkmark & X & X \\
 \hline
 $\slashed{P}_T + P_{T,L_1}$ and J B2B & 5 & X & \checkmark & X \\
 \hline
 $P_{T,L_0} + P_{T,L_1}$ and J B2B & 5 & X & X & \checkmark \\
 \hline
 1+ Secondary vertices & 6 & \checkmark & \checkmark & \checkmark \\
 \hline
 2 sub-jets & 7 & \checkmark & \checkmark & \checkmark \\
 \hline
 Simple Vertex cuts  & 8 & \checkmark & \checkmark & \checkmark \\
 \hline
 Machine Learning cuts & 9 & \checkmark & \checkmark & \checkmark \\
 \hline
\end{tabular}
\parbox{0.8\linewidth}{\caption{\label{tab:Cutflow}Cut flow for each channel.}}
\end{center}
\end{table}
\begin{enumerate}
\item MET 
    is reconstructed from the total sum of visible particles, with
    \begin{equation}
        |\eta|<4\;\;\mbox{\rm and}\;\;\;
        p_T>100\,\UMeV
    \end{equation}
    and it is particularly powerful in enhancing or suppressing events with (1L) and without (0L, 2L) decaying $W$ bosons.
\item to isolate leptons,  
    we demand that the total transverse energy of all particles in a cone of size $R_{\rm iso} = 0.2$ around the lepton direction is constrained by 5\% of the lepton transverse energy $E_{T,\ell}$, 
    \begin{equation}
        \sum_{\Delta R_{i}<R_{\rm iso}} E_{T,i}\;\leq\; 0.05\cdot E_{T, \ell}
    \label{eq:IsolatedLeptons}   
    \end{equation}
    We require the exact number of 0, 1, or 2 isolated leptons for the 0L, 1L, and 2L topologies.
\item 
    we demand the Higgs decay products to form a fat jet, defined by the anti-$k_{T}$ algorithm~\cite{Cacciari:2008gp} with  $R=1.0$ and $p_T > 250\,\UGeV$, and we require events to contain at least one such fat jet.  
\item 
    to identify the required single candidate fat jet, the highest-$p_T$ fat jet must contain at least three particles, but no isolated lepton, and its invariant mass must satisfy
    \begin{equation}
        75\,\UGeV<m_J<175\,\UGeV\,,
    \end{equation}
    cf.\ Fig.~\ref{fig:0LepFJMass} for an illustration that motivates our choice.
    \begin{figure}[tbp]
    \centering 
    \includegraphics[width=\linewidth]{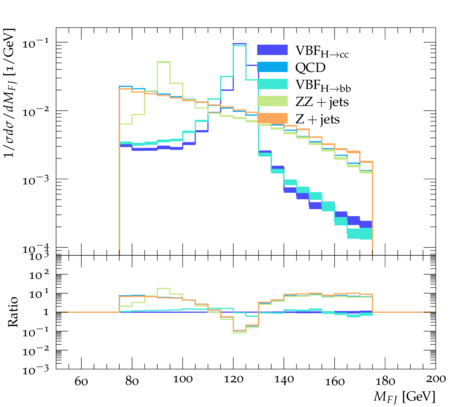}
    \parbox{0.8\linewidth}{\caption{\label{fig:0LepFJMass} Histogram of the reconstructed fat jet mass in the ``0L" channel. Shaded region indicates the error estimated from PDF4LHC scale variation convention and statistics.}}
    \end{figure}
\item 
    in addition, we place some cuts that uniquely identify specific signal topologies:
    \begin{itemize}
    \item 
        for the VBF (0L) topology, we require two forward anti-$k_T$ jets ($R=0.4$, $p_T>20\,\UGeV$), and with a minimal rapidity separation and combined invariant mass,
        \begin{equation}
            \Delta y_{jj}>2.5\;\;\;\mbox{\rm and}\;\;\;
            m_{jj}>400\,\UGeV\,.
        \end{equation}
    \item 
        for the two $VH$ topologies (1L and 2L), we demand that the combined momentum of the two isolated leptons (2L) or of the isolated lepton and MET (1L) is anti-parallel to the fat-jet momentum within a tolerance of $R = 0.4$.
    \end{itemize}
\item 
    The fat jet 
    must contain at least one reconstructed secondary vertex with at least two charged tracks.
    An adapted vertex fitter~\cite{Fruhwirth:1027031,ATL-PHYS-PUB-2017-011} performs a minimisation of weighted impact parameters, $d^{2}_i$ over points of closest approach of charged particles contained within the fat jet. 
    Vertices within 1\Ummeter of each other are considered unresolved and are merged.
\item 
    The fat jet must contain at least two sub-jets (anti-$k_T$: $\Delta R=0.4$, $p_T>20\,\UGeV$).
\item
    Further cuts are applied on the reconstructed primary vertex, namely mass flowing through it and the root mean square distance (RMSD) of particle tracks to the vertex, 
    \begin{equation}
        M_{v_p} < 1\UTeV\;\;\;\mbox{\rm and}\;\;\;
        RMSD_{v_p} < 3\Ummeter\,.
    \end{equation}
    which are determined empirically studying histograms of distributions over signal and background. 
\end{enumerate}
In Fig.~\ref{fig:0LepCutflow} we exhibit, as an example, the resulting cut flow for the 0L channel (vector boson fusion).

\section{Machine Learning improvements}
\subsection{ML ``booster"}
\noindent
In a second step, we boost the cut-based analysis through a set of multivariate neural networks (MVA) trained with TensorFlow~\cite{tensorflow2015-whitepaper} on $\sim$10,000 events of each background and signal processes that have survived the initial cuts:
\begin{enumerate}
\item 
    ``Observable": a dense fully connected network trained with global event and fat jet features.
    \begin{figure*}[tbp]
    \centering 
    \includegraphics[width=.3\textwidth]{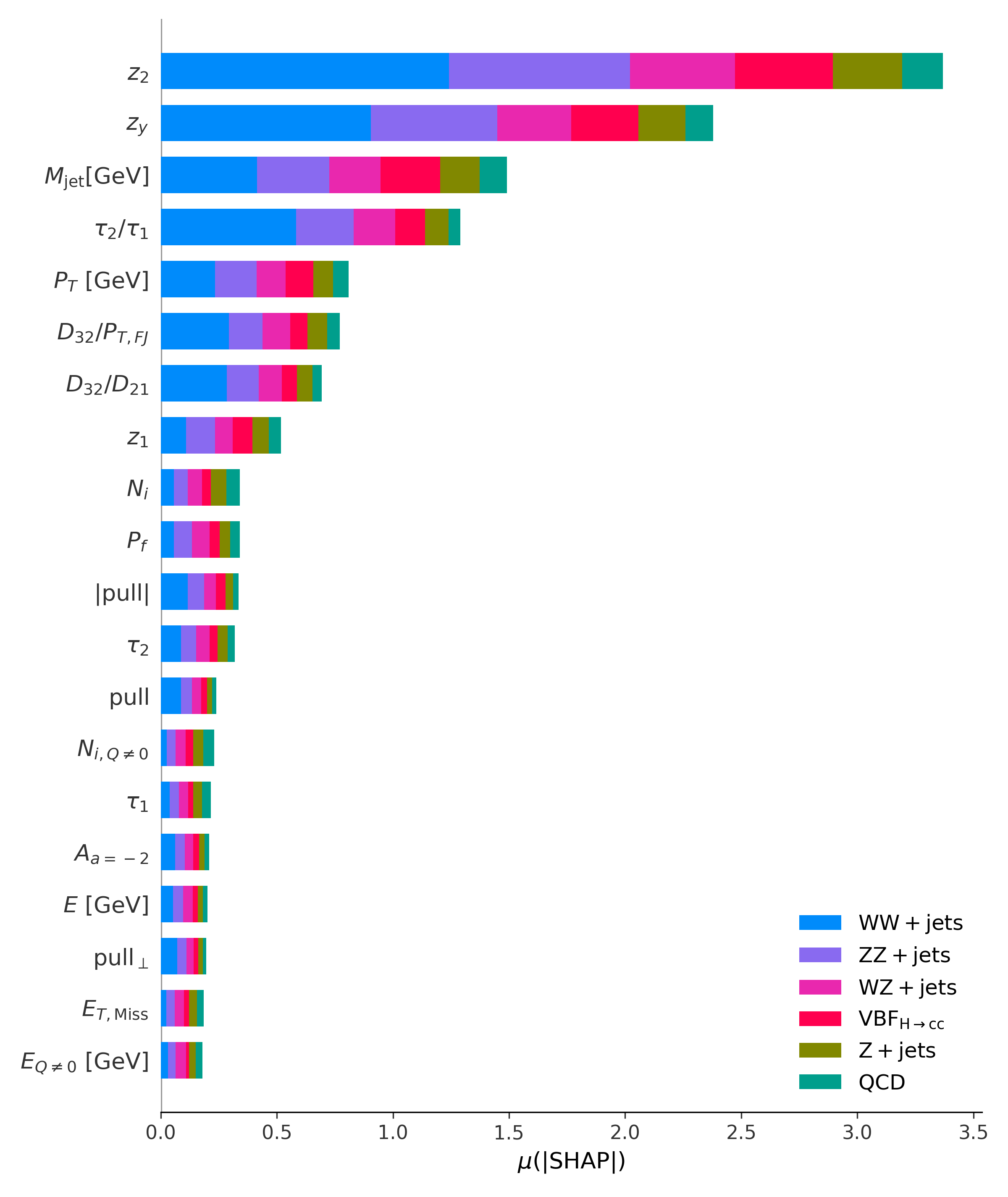}
    \hfill
    \includegraphics[width=.3\textwidth]{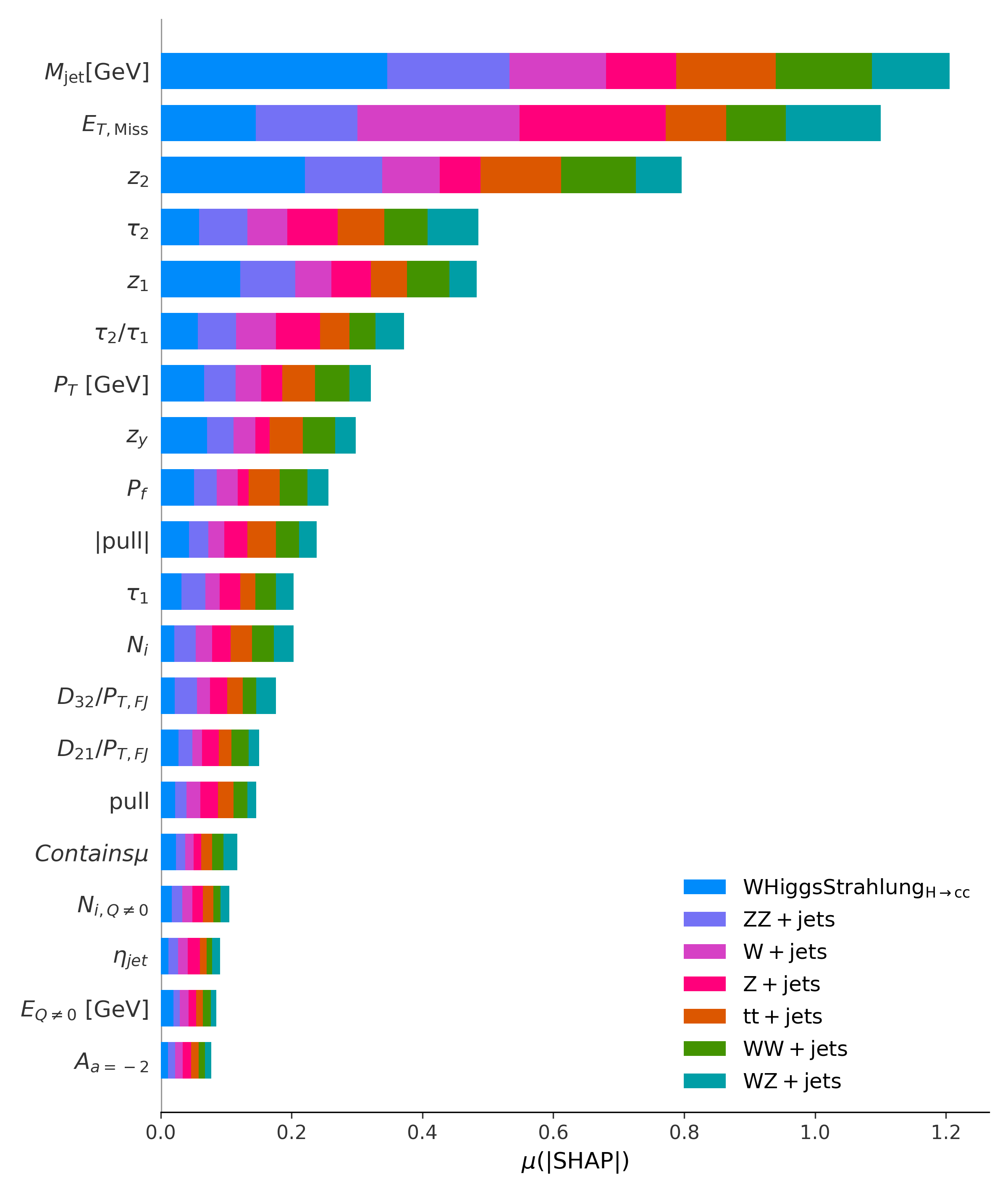}
    \hfill
    \includegraphics[width=.3\textwidth]{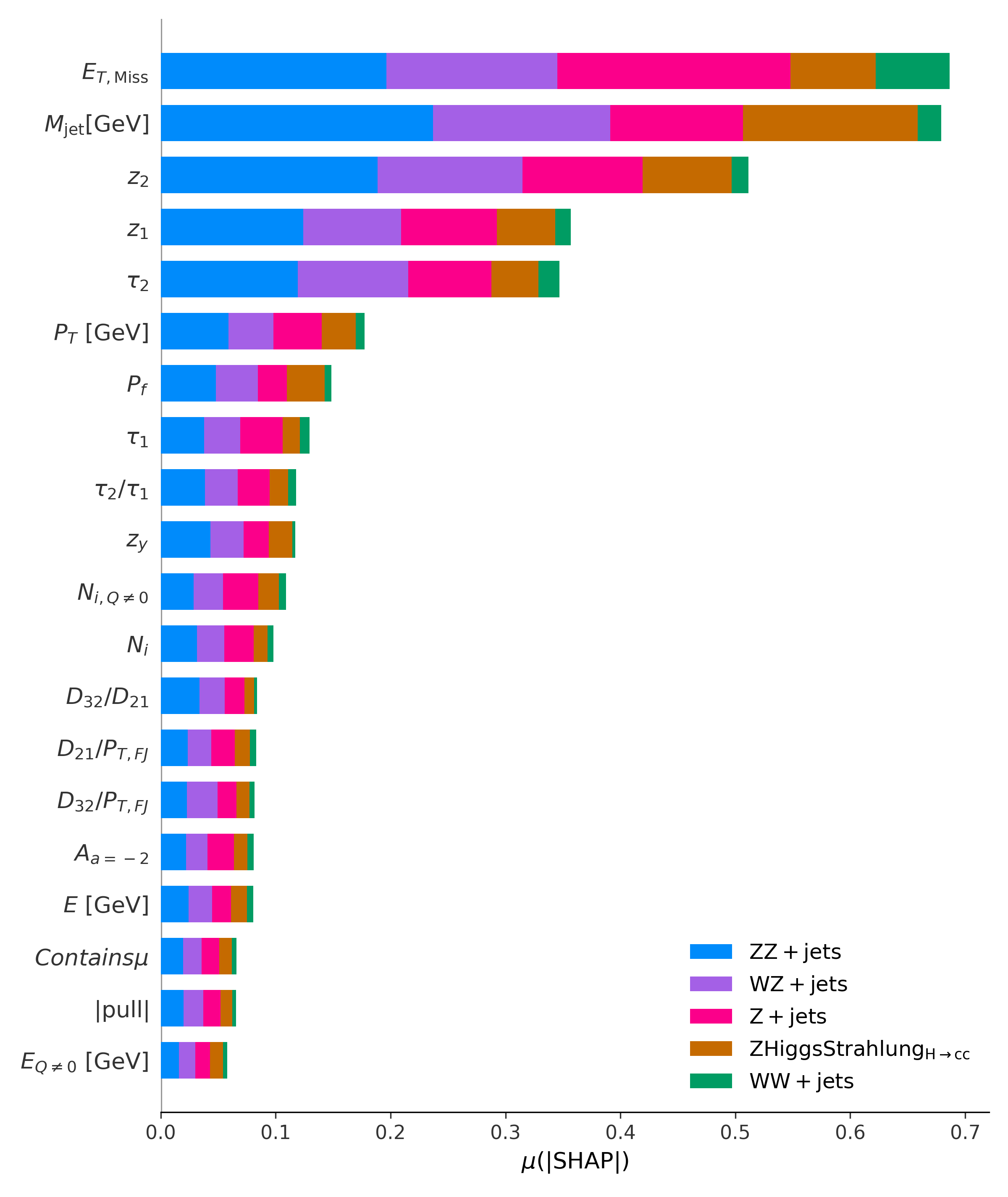}
    \parbox{0.8\textwidth}{\caption{\label{fig:ShapleyObservables} Absolute Shapley contributions to classification of each coloured class in each channel ``0L", ``1L", ``2L".}} 
    \end{figure*}
    
    \noindent
    A large selection of features are fed into the multivariate neural network.
    A number of global event and jet observables can be considered in order to best distinguish between the classes, and only the features with the strongest Shapley values~\cite{Shapley:1952} for the signal classes are used:
    \begin{itemize}
    \item jet mass: $m_{\textrm{J}}$,
    \item missing transverse energy: $\slashed{E}_T$,
    \item total perpendicular momentum in jet: $p_{\textrm{J} \perp}$,
        \begin{equation}
        p_{\textrm{jet} \perp} = \sqrt{p_x^2 + p_y^2}\quad \ni \quad \underline{p}_{\textrm{jet}} \cdot \hat{\underline{z}}=|\underline{p}_{\textrm{jet}}|
        \label{eq:jetperpmom}
        \end{equation},    
    \item 2-subjettiness: $\tau_{2}$~\cite{Davide:2018}, 
        The sub-jets required for the $N$-subjettiness calculation are clustered by the $\textrm{k}_{\textrm{t}}$ algorithm with $R = 1.0$ where clustering is stopped when exactly two sub-jets remain.
    \item sub-jet energy fraction: $z_{1}$ and $z_{2}$,
        \begin{equation}
        Z_{i} = \dfrac{E_\textrm{sub-jet,i} }{E_\textrm{J}}
        \label{eq:subjetEnergyFraction}
        \end{equation}    
    \item planar flow: $P_f$~\cite{2009PhRvD..79g4017A}.
    \end{itemize}
    \noindent
    These observables are standardised and normalised,
    \begin{equation}
    \begin{split}
    O_i^{\prime} =\;& \dfrac{O_i - \bar{O}}{\sigma_{O}} \\
    O_i^{\prime\prime} =\;& \dfrac{O_i^{\prime} - \textrm{min}(O)}{\textrm{max}(O) - \textrm{min}(O)} 
    \label{standarisedB}
    \end{split}
    \end{equation}
    This improves the gradient decent performance since it works more efficiently over variables with roughly equal ranges and magnitudes.
\item 
    ``Image": a dense fully connected convolutional neural network trained on rotated ``jet images".
    \begin{figure*}[!ht]
    \centering 
    \includegraphics[width=.45\textwidth]{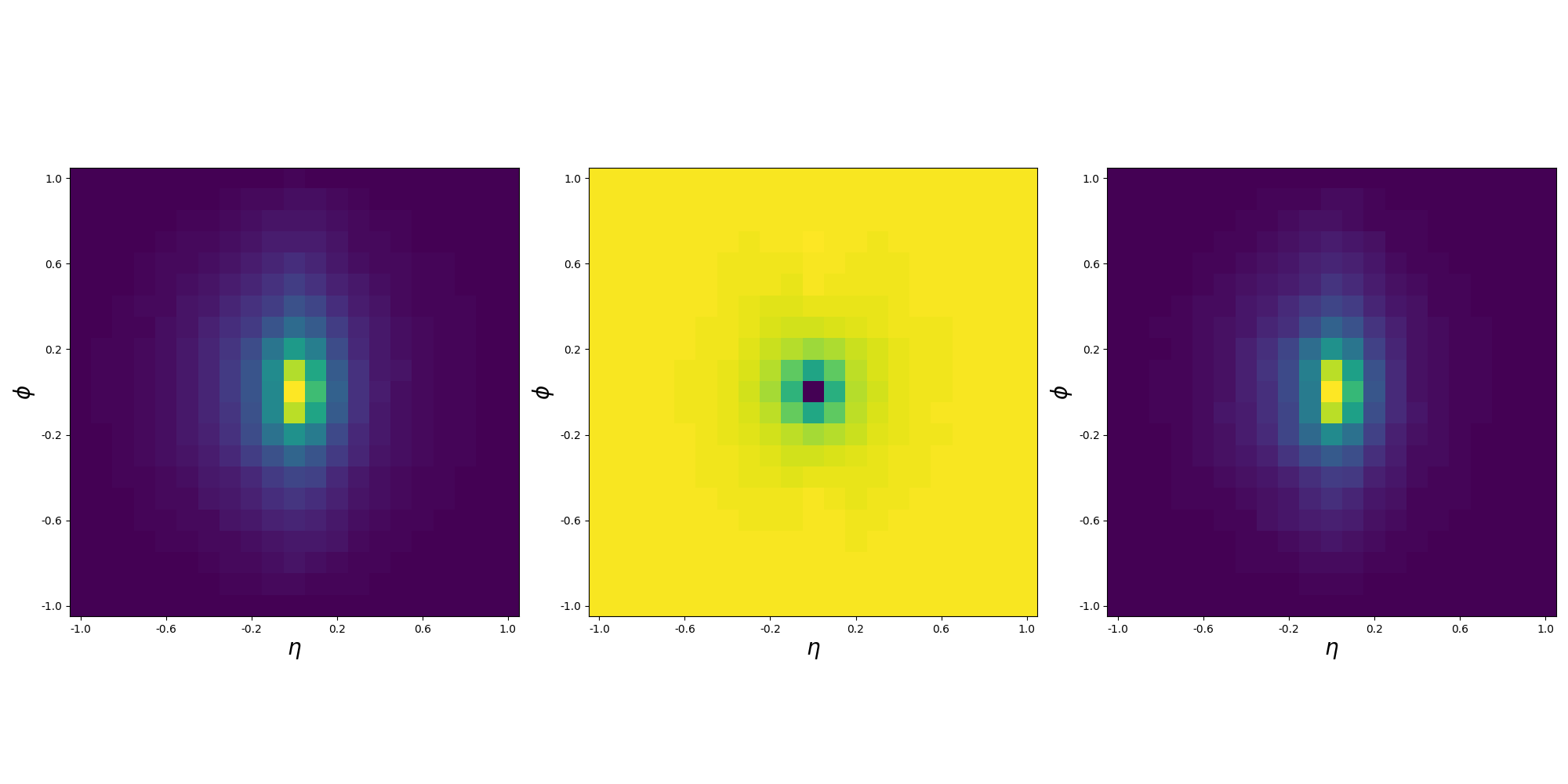}
    \hfill
    \includegraphics[width=.45\textwidth]{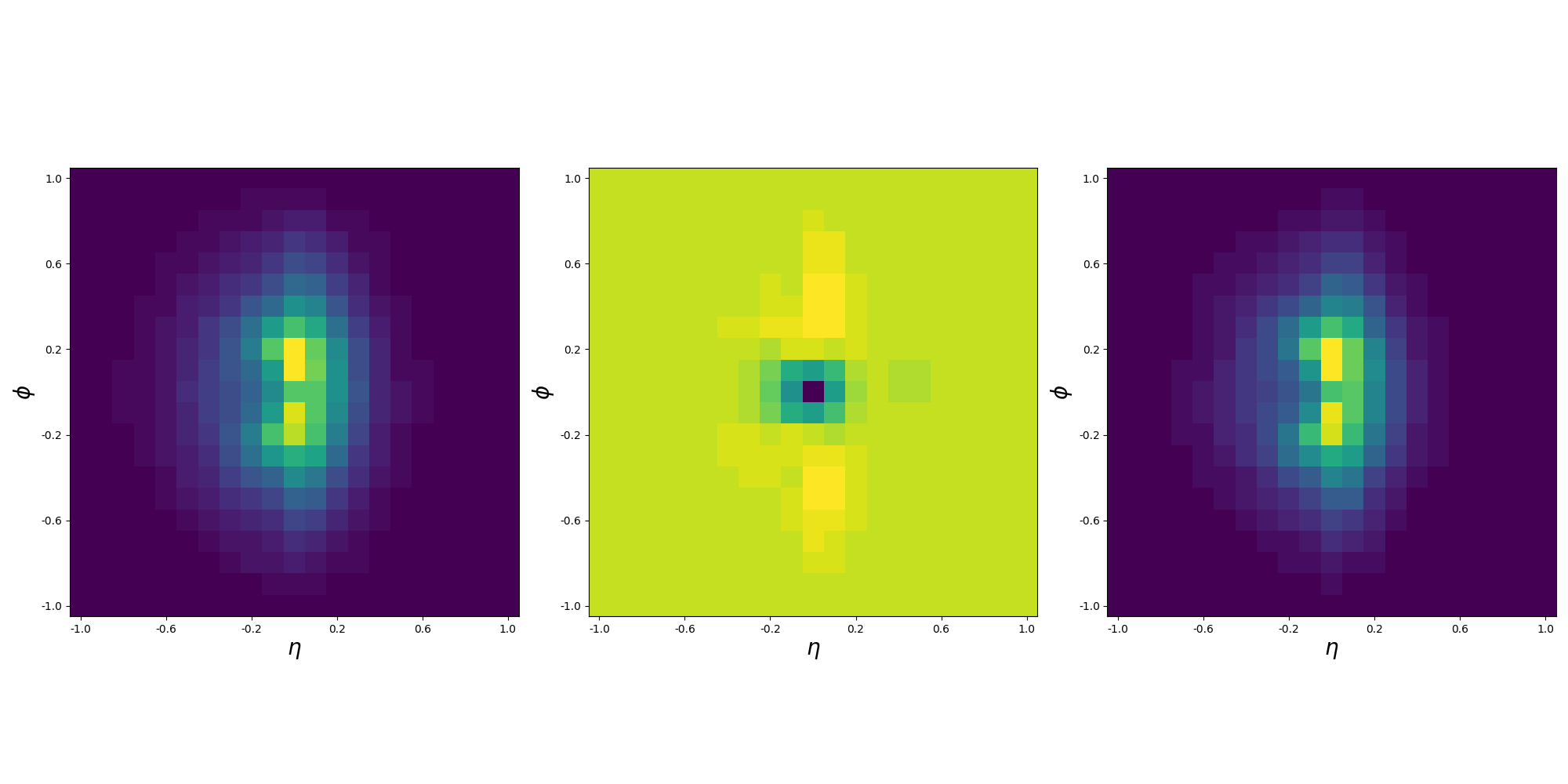}
    \parbox{0.8\textwidth}{\caption{\label{fig:ExampleImage} Example mean fat jet images for signal and background: QCD (left) vs.\ VBF($H_{\to c\bar{c}}$) (right).} }
    \end{figure*}
    We create ``2D calorimeter" images for the fat jet centred on its axis, and apply simple pre-processing steps to standardise the images using standard computer vision techniques,	
    \begin{enumerate}
	\item Center: Rotate the jet in ($\eta, \phi$) such that the jet axis lies at (0,0),
	\item Rotate: Rotate ($\eta, \phi$) such that any sub-jets align on the $\phi$-axis,
	\item Image set-up: An `image' which spans $\eta^{\prime}, \phi^{\prime} \in (-R,R)$ with 21x21 pixels, 
	\item Build: For each particle, $i$ in the jet add some variable $x$ in the bin ($\eta^{\prime}_i, \phi^{\prime}_i$),
	\item Scale: For each image scale such that $0 < I_{\eta^{\prime},\phi^{\prime}} < 255$,
    \end{enumerate}
    The algorithm above can be extended from grey-scale to a colour image, in this work we consider \textcolor{red}{R}\textcolor{green}{G}\textcolor{blue}{B} images where the pixel variable $x$ is,
    \begin{align*} 
        \textcolor{red}{R} &= -\Sigma_{\textrm{particles}} \log( E_i/E_{\textrm{jet}}) \\
        \textcolor{green}{G} &= -\Sigma_{\textrm{particles}} \log(p_{T,\textrm{jet}}) \\
        \textcolor{blue}{B} &= \Sigma_{\textrm{particles}, Q \neq 0}
    \end{align*}
    Jet images exemplified in Fig.~\ref{fig:ExampleImage} are fed into the CNN. 
    A pre-processing function centres and normalises the images to contain integer values between 0-255.  
    Similarly to the observables dataset this improves the gradient decent performance, allowing the network to be more sensitive to the full range of pixel values and enables faster learning during back-\-propagation. 
    The Image datasets also undergo some on-the-fly augmentation; once imported, there is a random chance of pixel shifting and randomised horizontal and vertical flipping. 
    These changes artificially generate more data and enforce discrete symmetries and any make the network robust to any centring issues encountered.
\item ``Flow": a dense fully connected recursive neural network trained on ordered particle level features within the fat jet.

    \noindent
    Particle level features are fed into a recursive neural network. 
    To provide the neural network with a structured sequence of particles, up to 10 fat jet constituents are ordered in energy.
    The particle level features are: 
    \begin{itemize}
	\item $\eta$ displacement w.r.t.\ fat jet axis: $\Delta \eta = \eta_p - \eta_{\textrm{J}}$,
	\item $\phi$ displacement w.r.t.\ fat jet axis: $\Delta \phi = \phi_p - \phi_{\textrm{J}}$,
	\item Perpendicular momentum: $\log (p_{T,p})$,
	\item Perpendicular momentum fraction: $\log (\dfrac{p_{T,p}}{p_{T,\textrm{J}}})$,
	\item Energy: $\log (E_{p})$,
	\item Energy fraction: $\log (\dfrac{E_{p}}{E_{\textrm{J}}})$,
	\item $R$ displacement w.r.t.\ fat jet axis, $\Delta R = R_p - R_{\textrm{J}}$,
    \end{itemize}
    Again we standardise these particle features across the training data set and all particles,
    \begin{equation}
        \begin{split}
        F_i^{\prime} =\;& \dfrac{F_i - \bar{F}}{\sigma_{F}}\\
        F_i^{\prime\prime} =\;& \dfrac{F_i^{\prime} - \textrm{min}(F)}{\textrm{max}(F) - \textrm{min}(F)}. 
        \end{split}
    \label{standarisedFB}
    \end{equation}
\end{enumerate}
\noindent
The structure of each of these parts are shown in Appendix~\ref{appendix2}, Fig.~\ref{fig:NNArchitectures}.

The neural networks are trained over many epochs with a test-\-validation split of 90\%-\-10\%, and the network with the highest validation accuracy from any epoch is kept. 
We note that at this stage there is confusion between the $H \to c\bar{c}$ and $H \to b\bar{b}$ class as this ML ``booster" makes no attempt to build a $b$ and $c$ jet classifier. 
The distinction between $H \to bb$ and $H \to cc$ is addressed with another network architecture, see below.
In each channel we report overall retention of signal excluding $H \to b\bar{b}$. 
The results can be summarised in Tab.~\ref{tab:MLresults} for $\epsilon_s$ and $\epsilon_b$ the signal acceptance and background rejection efficiencies (excluding $H \to b\bar{b}$).
\begin{table}[h]
\begin{center}
\begin{tabular}{ || c c || c c || } 
 \hline
 NN & ``\#L" & $\epsilon_s$ & $\epsilon_b$ \\ [0.5ex] 
 \hline\hline
 MVA & 0 & 82.0\% & 77.8\% \\ 
 \hline
 MVA & 1 & 82.4\% & 76.0\% \\ 
 \hline
 MVA & 2 & 69.0\% & 82.0\% \\ 
 \hline
\end{tabular}
\parbox{0.8\linewidth}{\caption{\label{tab:MLresults} Neural network efficiencies for each channel.}}
\end{center}
\end{table}
Only the events which are predicted to be the signal class for the appropriate channel are kept. 
The trained models are converted into a format suitable to run natively in \Rivet~\cite{Buckley:2010ar} using the Frugally Deep header library~\cite{FuggalyDeep}.

\subsection{ML charm vs.\ bottom discriminator}
Finally, we construct a neural network to discriminate the $H \to cc$ signal from the  $H \to bb$ background, based on the structure of the displaced vertices. 
There are many examples of superior classifiers for $c$ and $b$ jet classifiers used by the experiments, including MV2 and DL1~\cite{Aad_2019bbb}. 
However, here we construct our own multivariate neural network that uses primary and secondary vertex features to discriminate between fat jets with only light constituents, with $c$ hadrons, and with $b$ hadrons. 
This network has two inputs, 
\begin{enumerate}
\item ``vertex observable":
    a time-\-distributed fully connected network with input features describing up-to 5 reconstructed vertices, $V_i$ with 10 features: 
    \begin{itemize}
	    \item number of reconstructed vertices: $N_{V_i}$,
	    \item total number of tracks: $N_{p, V_i}$,
	    \item vertex invariant mass: $M_{V_i}$,
	    \item vertex Energy: $E_{V_i}$,
	    \item distance from primary vertex: $D(V_i, V_P)$,
	    \item transverse distance from primary vertex: $D_T(V_i, V_P)$,
	    \item RMSD of impact parameters: $\sqrt(\overline{d^{2}_i})$,
	    \item polar angle of vertex: $\theta_{V_i}$,
	    \item order of vertex: $\order_{V_i}$.
        \end{itemize}
    $\order_{V_i}$ is necessarily 0 for the primary vertex; then any vertices within a cone with opening angle $\theta = \pi/4$ are subsequently numbered in order of distance from the primary vertex. This provides the neural network with reinforcement of a natural ordering in displaced vertices in any event. 
\item ``vertex flow" 
    uses particle-level features of the 5 hardest particles of each vertex $V_i$ with the following inputs,
    \begin{itemize}
    \item longitudinal impact parameter: $d_{L,p}$,
    \item transverse impact parameter: $d_{T,p}$,
    \item energy fraction: $\log (\dfrac{E_{p}}{E_{\textrm{J}}})$,
    \item $\eta$ displacement w.r.t.\ fat jet axis: $\Delta \eta = \eta_p - \eta_{\textrm{J}}$,
    \item $\phi$ displacement w.r.t.\ fat jet axis: $\Delta \phi = \phi_p - \phi_{\textrm{J}}$,
    \item $R$ displacement w.r.t.\ fat jet axis: $\Delta R = R_p - R_{\textrm{J}}$.
    \end{itemize}
\end{enumerate}
All of these features are normalised over a weighted mean over all features for all classes, leading to the results summarised in Tab.~\ref{tab:MLresultsV}.
\begin{table}[h]
\begin{center}
 \begin{tabular}{ || c c || c c || } 
 \hline
 NN & ``\#L" & $\epsilon_s$ & $\epsilon_b$ \\ [0.5ex] 
 \hline\hline
 Vertex MVA & 0 & 54.8\% & 89.5\% \\ 
 \hline
 Vertex MVA & 1 & 42.2\% & 90.1\% \\ 
 \hline
 Vertex MVA & 2 & 46.6\% & 90.0\% \\ 
 \hline
\end{tabular}
\parbox{0.8\linewidth}{\caption{\label{tab:MLresultsV} Neural Network efficiencies for each channel.}}
\end{center}
\end{table}
The vertex booster network was also independently trained on a streamlined data set consisting of $H \to b\bar{b}$, $H \to c\bar{c}$ and QCD fat jets with a minimal cut flow for comparison to other analyses. 
We find $\epsilon_{H \to c\bar{c}} = 72\%$ and background rejection $\epsilon_b = 75\%$. 
Comparing directly with the JetFitterCharm Algorithm~\cite{ATL-PHYS-PUB-2015-001} and demanding similar signal efficiencies we obtain background rejection rates summarised in Tab.~\ref{tab:MLresultsJETTAGEFFSet}.
\begin{table}[h]
\begin{center}
 \begin{tabular}{ || c || c c c || } 
 \hline
  & $\epsilon_c$ & $1/\epsilon_b$ & $1/\epsilon_l$ \\ [0.5ex] 
 \hline\hline
 ``Loose"  & 0.95 & 1.65 & 1.03 \\ 
 \hline
 ``Medium"  & 0.21 & 13.2 & 149 \\ 
 \hline
\end{tabular}
\parbox{0.8\linewidth}{\caption{\label{tab:MLresultsJETTAGEFFSet} Summary of neural network efficiencies selected $\epsilon_c$ on the streamlined fat jet data set, resulting in $\epsilon_l$ light jet, $\epsilon_b$ bottom jet and $\epsilon_c$ charm jet efficiencies.}}
\end{center}
\end{table}
\noindent

\section{Results}
\subsection{Limitations of the $\kappa$ Framework}
To determine the 95\% confidence limit on the signal strength, $\mu_{0L}$, $\mu_{1L}$ and $\mu_{2L}$ we use a $\cls$~\cite{Read_2002} frequentist approach implemented in RooFit/RooStats~\cite{moneta2011roostats, Verkerke2003roostats} and treat \Sherpa as a standard model simulator. 
$\mu(\kappa, Br)$ is derived from equation~(\ref{eq:Mu2}) as
\begin{equation}
    \mu_{c} = 
        \dfrac{\kappa_c^2}{1+Br_{H\to c\bar{c}}^\textrm{SM}(\kappa_c^2-1)}
\label{eq:KctoMu}
\end{equation}

The \cls method uses a binned likelihood to determine the confidence limits on $\mu_i$. 
This likelihood incorporates  uncertainties due to statistics $\sigma_{N,x} = \sqrt{N}$, luminosity $\sigma_L = 2.5\%$~\cite{ATLAS-CONF-2019-021}, and scale variations $\sigma_\alpha$. 
Therefore,
\begin{equation}
\begin{split}
    \mathcal{L} & (\mu, \mathbf{s}, \mathbf{b}) =
    \mathcal{N}(L^{\prime}, L, \sigma_{L}) 
    \mathcal{N}(\alpha_s^{\prime}, \alpha_s, \sigma_{\alpha_s})
    \mathcal{N}(\alpha_b^{\prime}, \alpha_b, \sigma_{\alpha_b}) \cdot \\ 
    & \prod_x
    \mathcal{P}(b_x + s_x, L^{\prime}(\alpha_b^{\prime}b_x^{\prime} + \mu \alpha_s^{\prime}s_x^{\prime}))
    \mathcal{N}(b_x^{\prime}, b_x)
    \mathcal{N}(s_x^{\prime}, s_x)
\end{split}
\label{eq:likelyhood}
\end{equation}

\begin{figure}[tbp]
\centering 
\includegraphics[width=\linewidth]{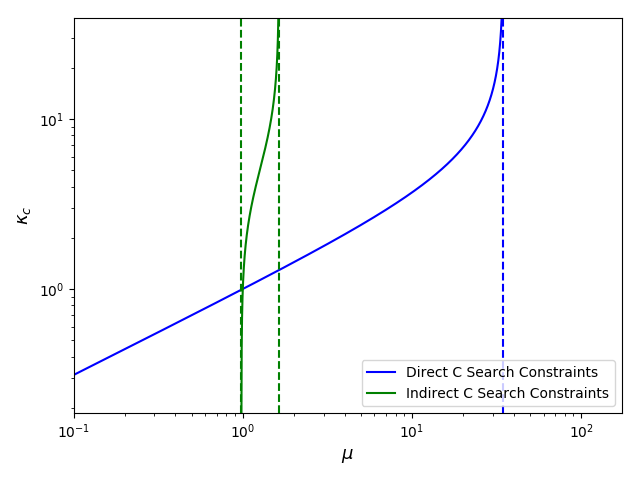}
\parbox{0.8\linewidth}{\caption{\label{fig:KctoMu} $\kappa(\mu_i)$ plotted in three instances. Blue: direct $H \to c\bar{c}$ defined in equation~(\ref{eq:KctoMu}). Green: The indirect measurement defined in Eq.~(\ref{eq:KcKbtoMu}).}}
\end{figure}

The uncertainties are parameterised with a Gaussian smearing over our expected values, and the priors $\mathcal{N}$ and $\mathcal{P}$ are Gaussian and Poisson distributions, respectively. 
Profiling the likelihood function for each channel determines the confidence limits as a function of $\mu_i$, with $i=$ ``0L", ``1L" or ``2L". 
These independent channels are combined into one confidence limit which could be inverted to a confidence limit on $\kappa_c$. 
However, inverting equation~(\ref{eq:KctoMu}) is not well defined for $\mu_{c} > 1/Br_{H \to cc}$, as illustrated in Fig.~\ref{fig:KctoMu}. 
As we explore this region of $\mu$-values with our analysis, and in order to avoid counter-intuitive results for $\kappa_c$, we only quote projections for limits on $\mu$.
We also suggest an indirect measurement, in which $H \to cc$ and $H \to bb$ are combined to the signal class and modified together with $\mu_{cb}$. 
This extension transforms Eq.~(\ref{eq:KctoMu}) into,
\begin{equation}
\begin{split}
    \mu_{cb} =  & \dfrac{
        \kappa_c^2 Br_{H \to cc}^{\textrm{SM}} +  \kappa_b^2 Br_{H \to bb}^{\textrm{SM}}
    }{
        ( Br_{H \to cc}^{\textrm{SM}} +Br_{H \to bb}^{\textrm{SM}} )
    } \cdot \\
    & \dfrac{1}{(1 + Br_{H \to bb}^{\textrm{SM}}(\kappa_b^2-1) + Br_{H \to cc}^{\textrm{SM}}(\kappa_c^2-1 ))}
\end{split}
\label{eq:KcKbtoMu}
\end{equation}
and the limitation on $\kappa$ becomes $\mu_{cb} > 1/(Br_{H \to cc}+Br_{H \to bb})$.

\subsection{$\mu$ results}
In each channel we consider the primary contributions of uncertainties in the determination of $\mu$. 
This can be done by fixing nuisance parameters in turn and varying each quantity by its pre-fit (initial uncertainties fed into likelihood) and post-fit (post fitted values from maximization of profiled likelihood) values. 
\noindent
\begin{figure*}[tbp]
\centering 
\includegraphics[width=.3\textwidth]{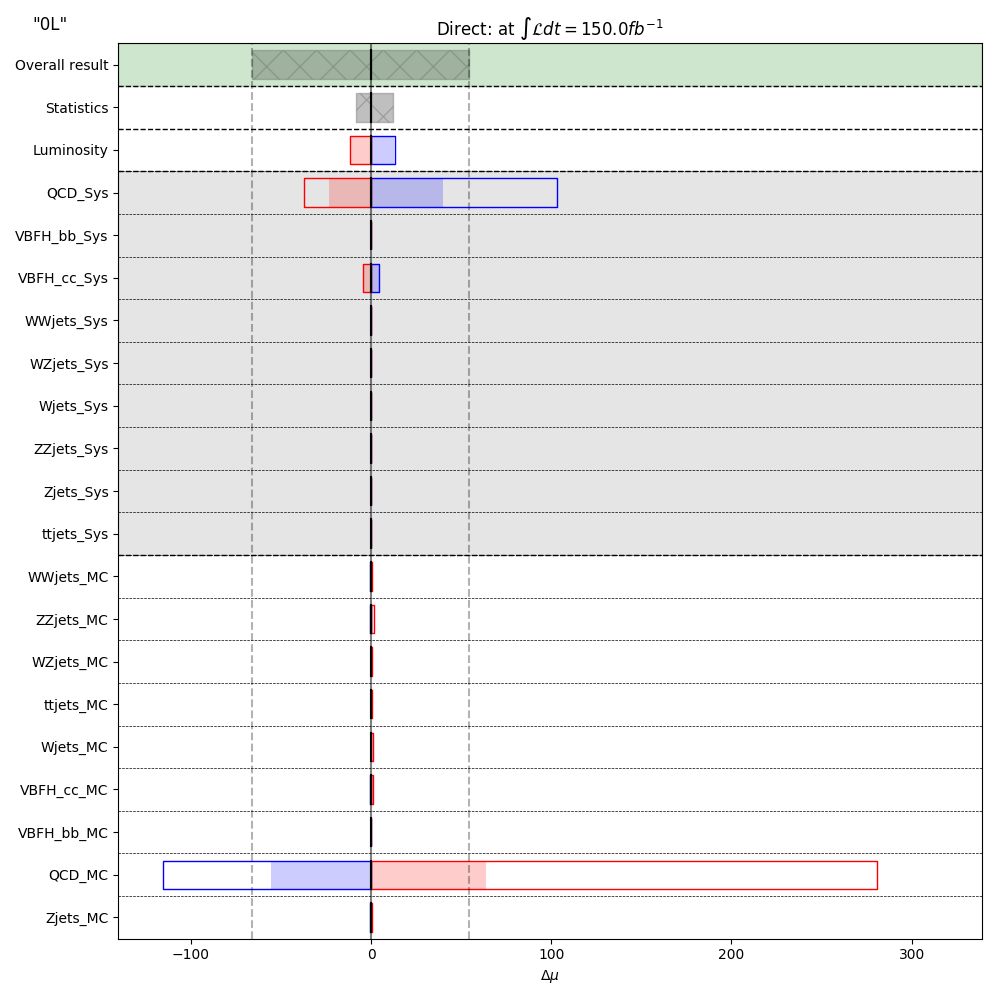}
\hfill
\includegraphics[width=.3\textwidth]{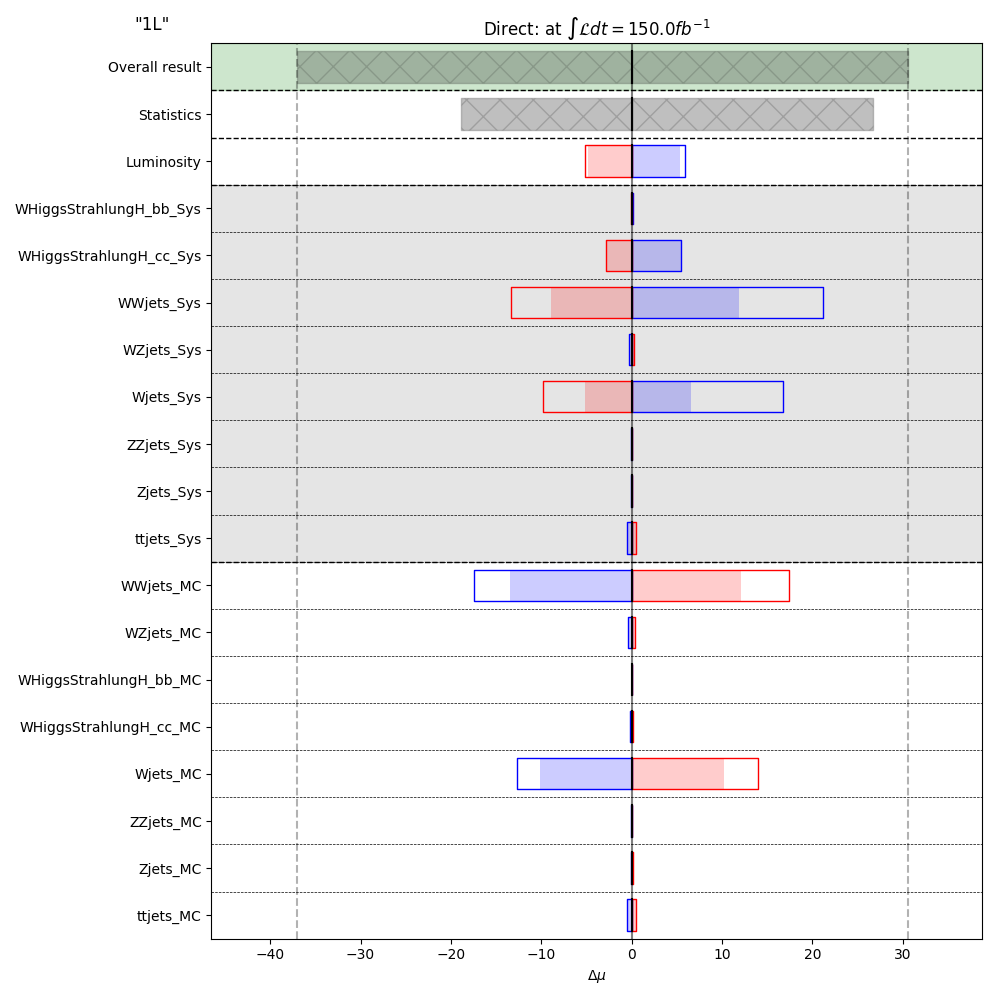}
\hfill
\includegraphics[width=.3\textwidth]{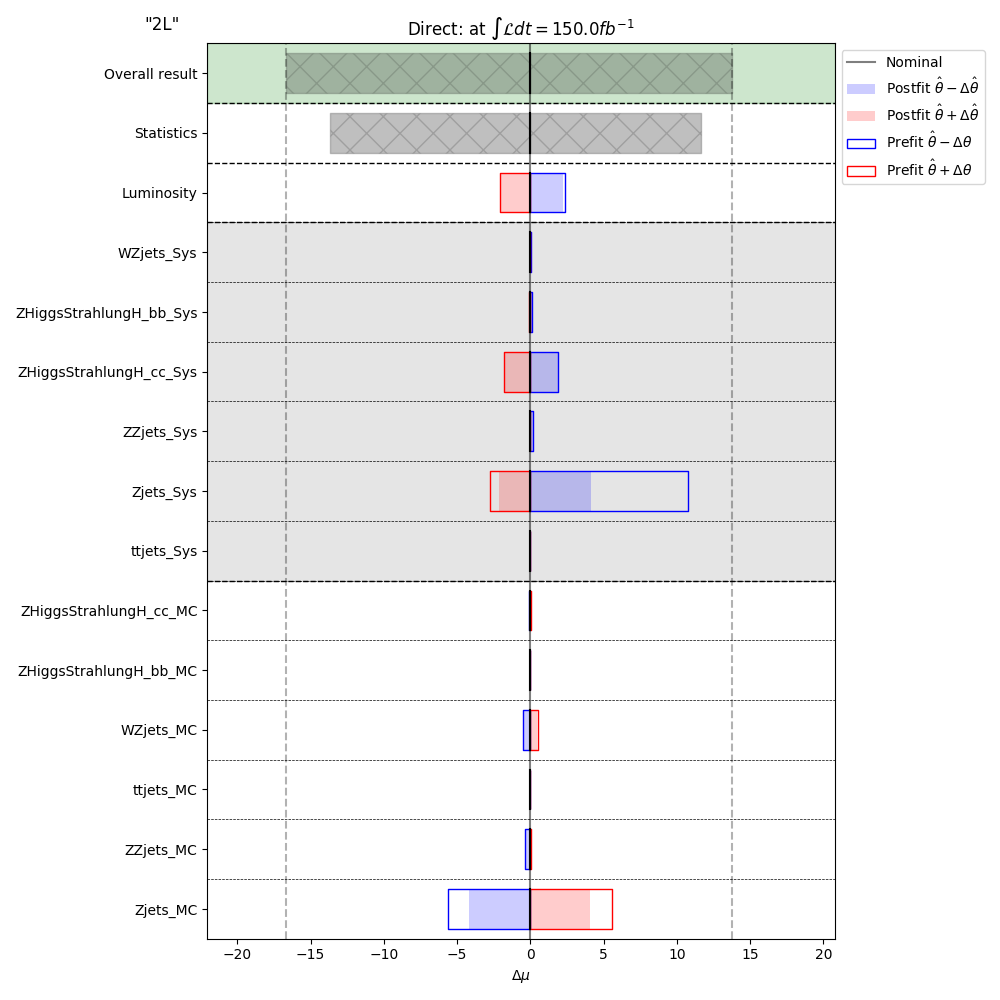}
\parbox{0.8\textwidth}{\caption{\label{fig:MarginalisedErrors} 
    Uncertainty contributions to the 95\% confidence limit of signal strength $\mu$ for $\mathcal{H}_{M_J}$ for the jet mass distribution. Uncertainties are shown for total statistics, luminosity, systematic uncertainty and Monte Carlo simulation uncertainty over all signal and background classes for each channel, ``0L", ``1L", ``2L", and compared with the total uncertainty. Pre-fit uncertainties shown in the outlined bars and post-fit uncertainties by the filled in bars.} }
\end{figure*}
In Fig.~\ref{fig:MarginalisedErrors} we exhibit four classes of uncertainties, 
\begin{enumerate}
    \item Statistical
    \item Luminosity
    \item Systematics (``Sys")
    \item Monte-carlo (``MC")
\end{enumerate}
Statistical uncertainties occur from the total counts in each bin in the likelihood function. The systematics from the 7-point envelope function in scale variations of $f_F, f_R$ and lastly the Monte Carlo uncertainty from \Sherpa.  
From this we can read off which backgrounds have the largest impact on the precise determination of $\mu$: QCD, $W$+jets ($W$) production and $Z$+jets ($Z$) production processes for the ``0L", ``1L" and ``2L" channels, respectively. 
Profiling nuisance parameters to calculate their marginal error allows the determination of correlations between other parameters. 
It is worth stressing that measurements in the ``2L" channel, while being the most sensitive channel in this analysis, are dominated by the statistics from the limited cross-section of the signal process and lower luminosity. 

We explored four ways for the $\mu$ extraction, direct (1), indirect (2), direct with $b\leftrightarrow c$ discrimination (3), and indirect with $b\leftrightarrow c$ discrimination (4).
We summarise the interplay of the different analysis steps and the variation of $\mu_c$ and $\mu_{cb}$ in Tab.~\ref{tab:methods}.
\begin{table}[ht]
\begin{center}
 \begin{tabular}{ || c | c | c | c | c || } 
 \hline
  & cut flow & $\textrm{ML}_{booster}$ & $\textrm{ML}_{b \leftrightarrow c}$ & $\kappa$ dependence \\ [0.5ex] 
 \hline\hline
 1 & \checkmark  & \checkmark  & X & $\mu_c(\kappa_c)$ \\ 
 \hline 
 2 & \checkmark  & \checkmark  & X & $\mu_{cb}(\kappa_c, \kappa_b)$ \\ 
 \hline 
 3 & \checkmark  & \checkmark  & \checkmark & $\mu_c(\kappa_c)$ \\ 
 \hline 
 4 & \checkmark  & \checkmark  & \checkmark & $\mu_{cb}(\kappa_c, \kappa_b)$ \\ 
 \hline 
\end{tabular}
\parbox{0.8\linewidth}{\caption{\label{tab:methods} Summary of the cut flows and neural networks architectures used in each of the methods and which of the $\kappa_{c}, \kappa_{b}$ that are allowed to vary. 
Here $\textrm{ML}_{booster}$ refers to the ML ``booster" network and $\textrm{ML}_{b \leftrightarrow c}$ the ML $b \leftrightarrow c$ discriminator network.}}
\end{center}
\end{table}
Fitting to distributions of various observables or to pairs of observables yields constraints of the the signal strengths $\mu_c$.  
The following obervables showed the best discriminating power:
\begin{itemize}
    \item planar flow: $P_f$,
    \item 2-subjettiness: $\tau_2$,
    \item boosted sub-jet separation angle: $\theta_{j1, j2}$,
    \item fat jet mass: $M_{\textrm{J}}$,
    \item sub-jet energy fraction: $Z_1$.
\end{itemize}
In Appendix~\ref{appendix1} we show results for the four methods over all distributions for an integrated luminosity of $150\,\mbox{\rm fb}^{-1}$ in Fig.~\ref{fig:AllCompares150}, and for $3\,\mbox{\rm ab}^{-1}$ in Fig.~\ref{fig:AllCompares3000}. 
The figures exhibit the 95\% confidence limits of all considered 1 dimensional fits and 2 dimensional fits with the $1\sigma$ and $2\sigma$ uncertainty bands. 

\begin{table*}[ht]
\begin{center}
 \begin{tabular}{ || c || c | c | c || } 
 \hline
 Search & $\mu$ w/ $\mathcal{H}_{M_J}$ & $\mu$ w/ $\mathcal{H}_{\textrm{best}}$ & $\mathcal{H}_{\textrm{best}}$ \\ [0.5ex] 
 \hline \hline & & & \\ [-1em] 
 $\int \mathcal{L} dt = 150 fb^{-1}$ & & & \\  [0.5ex] 
 \hline & & & \\ [-1em]  
 Direct, $\mu_c$ & $53.7^{+22.7}_{-15.9}$ & $42.7^{+17.4}_{-12.4}$ & $\mathcal{H}(M_J, P_f)$ \\  [0.5ex] 
 \hline & & & \\ [-1em] 
 Indirect, $\mu_{cb}$ & $4.0^{+1.3}_{-0.9}$ & $3.1^{+1.0}_{-0.7}$ & $\mathcal{H}(M_J, \theta_p)$  \\  [0.5ex] 
 \hline & & & \\ [-1em] 
 Direct, $\mu_c$ with $b\leftrightarrow c$ discrimination & $48.1^{+19.2}_{-13.8}$ & $8.0^{+3.6}_{-2.3}$ & $\mathcal{H}(Z_1, P_f)$  \\  [0.5ex] 
 \hline & & & \\ [-1em] 
 Indirect, $\mu_{cb}$ with $b\leftrightarrow c$ discrimination & $4.7^{+1.6}_{-1.1}$ & $2.0^{+0.6}_{-0.4}$ & $\mathcal{H}(Z_1, P_f)$   \\  [0.5ex] 
 \hline\hline & & & \\ [-1em] 
 $\int \mathcal{L} dt = 3000 fb^{-1}$ & & & \\  [0.5ex] 
 \hline & & & \\ [-1em] 
 Direct, $\mu_c$ & $35.5^{+14.0}_{-10.3}$ & $12.1^{+5.1}_{-3.4}$ & $\mathcal{H}(M_J,P_f)$  \\  [0.5ex] 
 \hline & & & \\ [-1em] 
 Indirect, $\mu_{cb}$ & $3.0^{+0.8}_{-0.6}$ & $1.5^{+0.2}_{-0.2}$ & $\mathcal{H}(M_J, \theta_p)$ \\  [0.5ex] 
  \hline & & & \\ [-1em] 
 Direct, $\mu_c$ with $b\leftrightarrow c$ discrimination & $33.9^{+13.2}_{-8.9}$ & $2.1^{+0.6}_{-0.4}$ & $\mathcal{H}(Z_1, P_f)$ \\  [0.5ex] 
 \hline & & & \\ [-1em] 
 Indirect, $\mu_{cb}$ with $b\leftrightarrow c$ discrimination  & $4.0^{+1.2}_{-0.8}$ & $1.1^{+0.1}_{-0.1}$ & $\mathcal{H}(Z_1, P_f)$ \\ [0.5ex] 
 \hline
\end{tabular}
\parbox{0.8\linewidth}{\caption{\label{tab:KappaResults} Combined $\mu$ bounds at 95\% confidence for 4 methods for the $\mathcal{H}_{M_J}$ and $\mathcal{H}_{\textrm{best}}$ distributions.
We compare results obtained from the fat jet distribution ($\mathcal{H}_{M_J}$) -- the method of choice for the experimental analyses so far -- with results we obtained from a best combination of two observables, $\mathcal{H}_{\rm best}$ and indicate them in the last column.}}
\end{center}
\end{table*}

One way, in which the impact dominant statistical uncertainties can be counteracted, is by performing a two-\-dimensional fit and constraining the sum of bins on each axis to one another and therefore their uncertainty to one another. 
We summarize the $\mu$ limits we obtain in Tab.~\ref{tab:KappaResults} for the standard choice $\mathcal{H}(M_J)$, used by the experimental analyses so far, and our most powerful combination of distributions, $\mathcal{H}_{\textrm{best}}$. 
Due to the low branching fraction of $H \to c{\bar{c}}$ we have a very low signal count at low luminosity, and moving forward into the high-\-luminosity phase of the LHC with around $3000\textrm{fb}^{-1}$ we will be less limited by statistics. 
This impacts on the efficiency of the ML ``booster" cuts, which greatly improve our confidence limit by a factor of 2 over the initial cuts at $150\textrm{fb}^{-1}$.

The two-\-dimensional fitting technique leads to an improvement in the obtained confidence limits by on average a factor of $2$ over their one-\-dimensional counterparts. 
We also see that while distributions involving $M_J$ provide good fits, other choices of observable work just as well or even better hinting at possible new avenues of exploration. 
The ML $b \leftrightarrow c$ discriminator network provides an improvement to the value of the confidence limit of a factor $2$ in the direct case but only $1.1$ in the indirect case. 
Our best fit result is $\mu_c \le 8.0^{+3.6}_{-2.3}$ at $150\textrm{fb}^{-1}$  ($\kappa_c \le 3.18^{+0.94}_{-0.60}$) at the 95\% confidence limit. This result is compatible to SM within 4.0 standard deviations and is competitive with current ATLAS and CMS values of $31^{+12}_{-8}$~\cite{ATLAS:2021zwx} and $37^{+16}_{-11}$~\cite{CMS:2019tbh}. 
These results may have scope for enhancement by considering a wider range of features and multi-dimensional fitting rather than the using the ``standard" choice as we have demonstrated in our findings. 

Moving into the high--\-luminosity regime we see a again an enhancement in the benefit from 2D fits by a factor of about $2$ over 1D fits. At $3 \rm{ab}^{-1}$ our best fit $\mu$ values tighten and the limits are now resolvable under the $\kappa$ framework. 
The direct measurement provides the best expected limit of $\kappa_c \le 1.47^{+0.21}_{-0.16}$ at the 95\% confidence limit. 

 


\section{Conclusions}
We studied prospects for a determination of the charm Yukawa coupling or its constraints with present and future LHC data. 
We considered the production of the Higgs boson in Higgstrahlung and weak boson fusion processes, leading to final states with 0, 1, or 2 leptons, and the subsequent decay of the Higgs boson into a fat jet.
We augmented a simple cut-based strategy with a multi-variate ``booster" step and showed that this enhances the sensitivity of the analysis.
We also investigated the impact of a neural-network based discriminator for fat jets containing only light partons, charm or bottom quarks and found a non-negligible impact.
As a by-product we suggested an indirect measurement strategy where the branching ratio of a Higgs boson into heavy quarks -- charm or bottom -- is used in conjunction with the known value of the bottom Yukawa coupling to infer the charm Yukawa coupling.
The signal strength in any case is extracted from fits to observable distributions, and we found that the fat jet mass is not necessarily the best-suited observable.
We also found that two-dimensional fits further boost the sensitivity by about factors of two to four compared to fits to a single observable, motivating further investigations.
\section*{Acknowledgements}
We are indebted to Marumi Kado and Francesco Di Bello for enlightening discussions and important feedback throughout the project, and we wish to thank the other members of the \Sherpa team for ongoing collaboration.

\noindent
This work was supported by the UK Science and Technology Facilities Council (STFC) under grant \linebreak ST/P001246/1. 
FK acknowledges support from the European Union's Horizon 2020 research and innovation programme as part of the Marie Sklodowska-Curie Innovative Training Network MCnetITN3 (grant agreement no. 722104), and support by the Wolfson Foundation and the Royal Society under award RSWF\textbackslash{}R1\textbackslash{}191029.

\FloatBarrier
\bibliographystyle{elsarticle-num}

\clearpage
\onecolumn
\begin{appendix}
\section{Likelihood distributions}
\label{appendix1}
\begin{figure}[!htb]
\centering 
\begin{tabular}{cc}
\includegraphics[width=.35\textwidth]{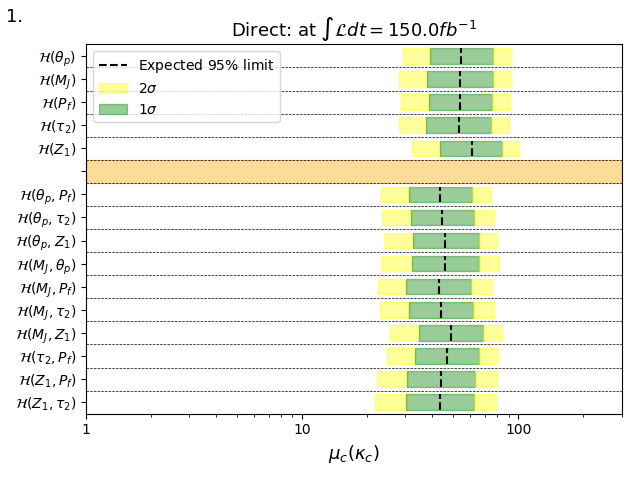} &
\includegraphics[width=.35\textwidth]{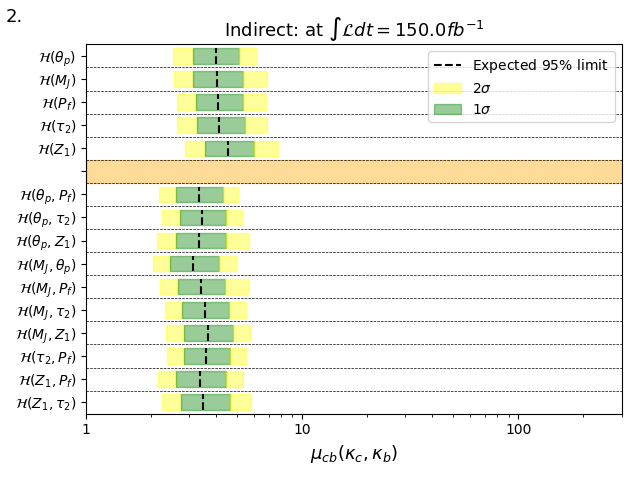}
\\
\includegraphics[width=.35\textwidth]{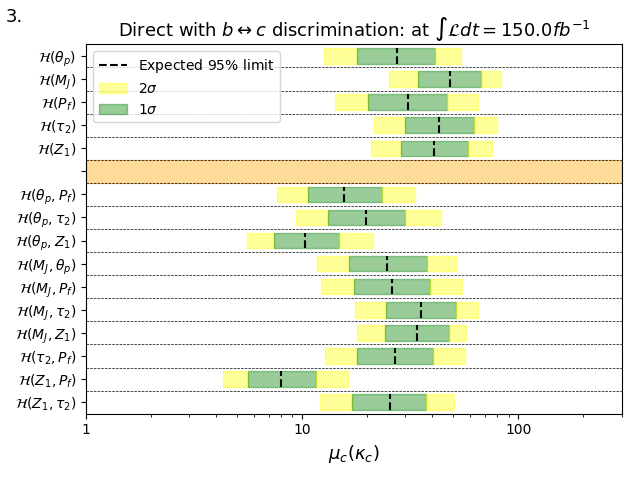}
&
\includegraphics[width=.35\textwidth]{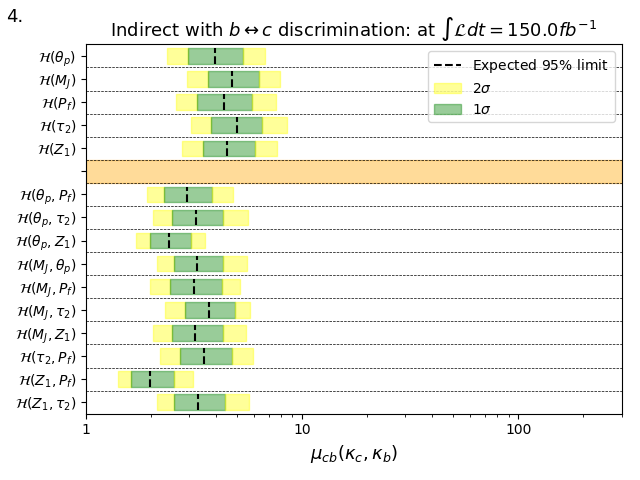}
\end{tabular}
\caption{\label{fig:AllCompares150} Comparison over binned likelihood distributions showing the 95\% confidence limit on the signal strength $\mu$.}
\end{figure}

\begin{figure}[!htb]
\centering 
\begin{tabular}{cc}
\includegraphics[width=.35\textwidth]{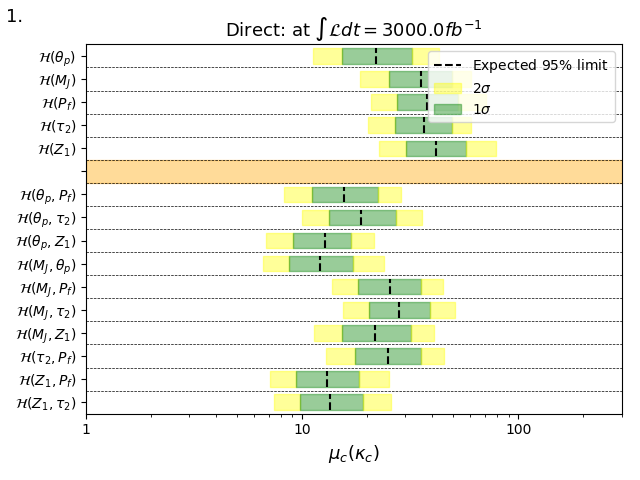} &
\includegraphics[width=.35\textwidth]{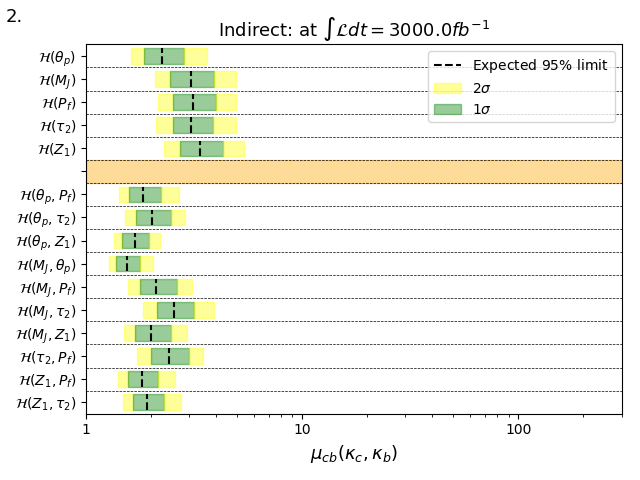}
\\
\includegraphics[width=.35\textwidth]{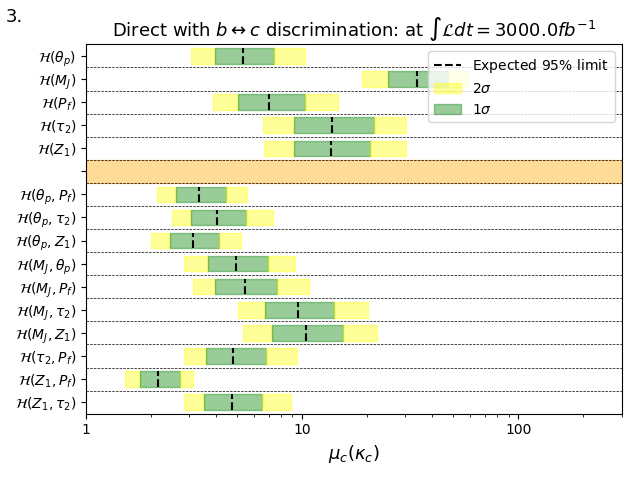}
&
\includegraphics[width=.35\textwidth]{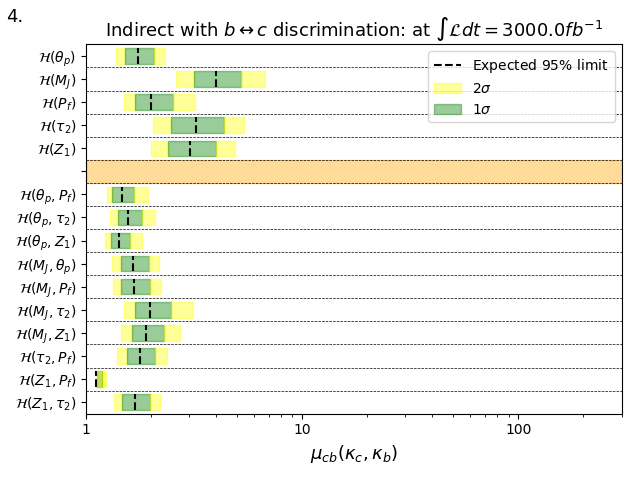}
\end{tabular}
\caption{\label{fig:AllCompares3000} Comparison over binned likelihood distributions showing the 95\% confidence limit on the signal strength $\mu$.}
\end{figure}

\section{Neural Network architectures}
\label{appendix2}
\begin{figure}[!htb]
\centering 
\begin{tabular}{c}
\includegraphics[width=.95\textwidth]{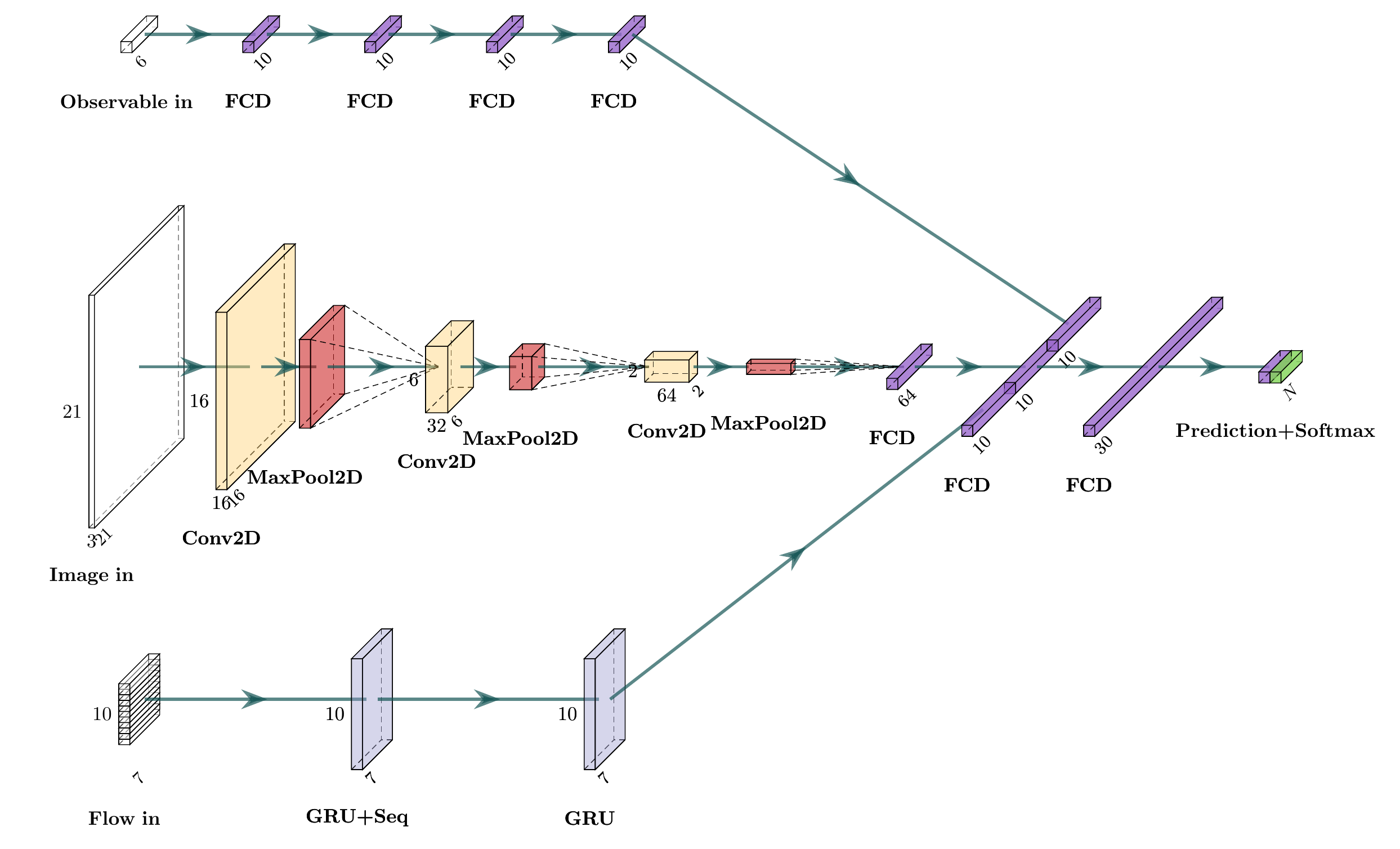}
\\
\hline \\
\includegraphics[width=.95\textwidth]{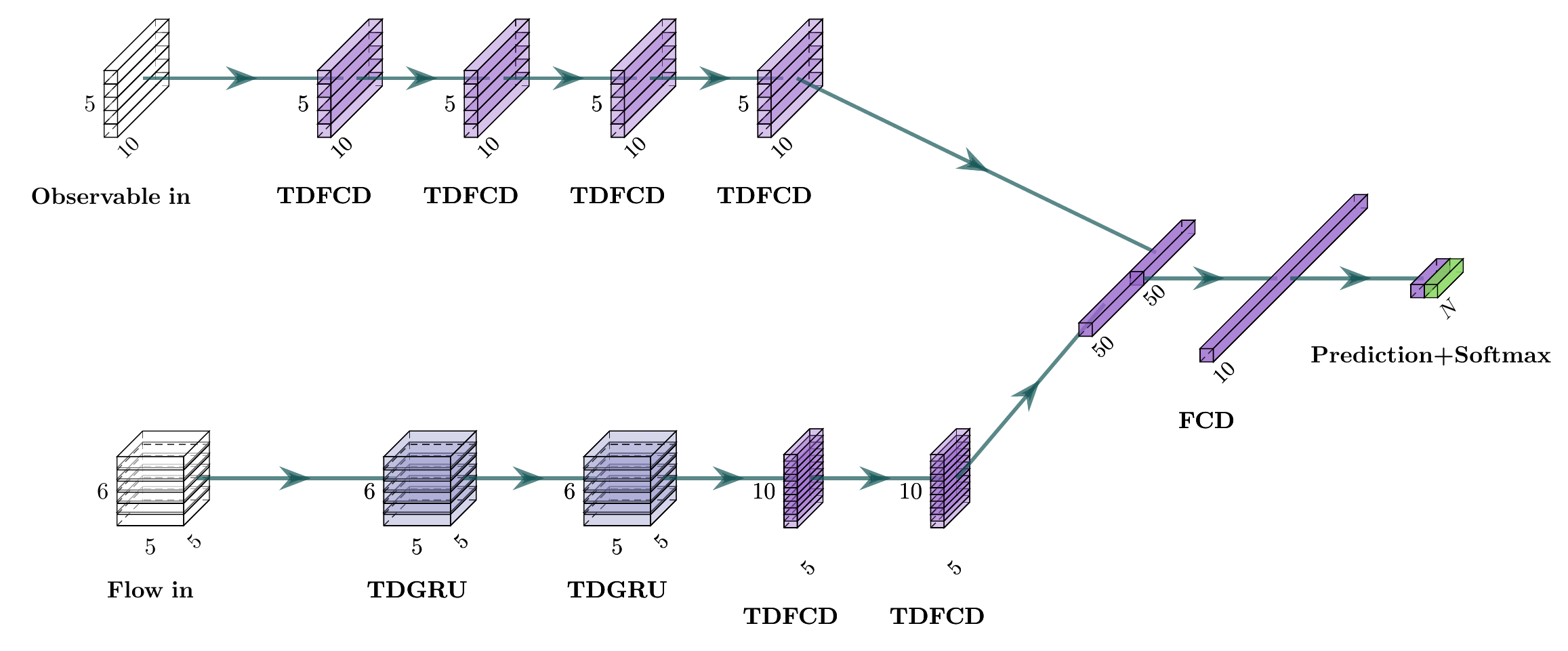}
\end{tabular}
\caption{\label{fig:NNArchitectures}  The neural network architectures for the $\textrm{ML}_{booster}$ (top) and $\textrm{ML}_{b \leftrightarrow c}$ (bottom). The abbreviation layer names are; \textbf{FCD}: Fully connected dense layer, \textbf{Conv2D}: A two dimensional convolutional layer, \textbf{MaxPool2D}: A two dimensional maximum pooling layer, \textbf{GRU+Seq}: A Gated recurrent unit which returns output from each unit not only the last, \textbf{GRU}: A gated recurrent unit, \textbf{TDFCD}: Time distributed fully connected dense layer and lastly \textbf{TDFCD}: Time distributed gated recurrent unit. Drop out layers used in training are omitted from these diagrams.}
\end{figure}
\end{appendix}
\twocolumn
\end{document}